\pdfoutput=1
\documentclass{nature_mod}
\usepackage{amsmath}
\usepackage{amssymb}
\usepackage{pdfpages}

\title{Supplementary information: Quantum fluctuations can promote or inhibit glass formation}
\author{Thomas E. Markland$^{1}$, Joseph A. Morrone$^1$, B. J. Berne$^1$, Kunimasa Miyazaki$^2$, Eran Rabani$^3$ \& David R. Reichman$^1$}

\begin{document}



# Quantum fluctuations can promote or inhibit glass formation

Thomas E. Markland,[1] Joseph A. Morrone,[1] B. J. Berne,[1,*] Kunimasa Miyazaki,[2] Eran Rabani,[3,*] David R. Reichman[1,*]

[1]*Department of Chemistry, Columbia University, 3000 Broadway, New York, New York, 10027, United States*

[2]*Institute of Physics, University of Tsukuba, Tsukuba, Japan*

[3]*School of Chemistry, The Sackler Faculty of Exact Sciences, Tel Aviv University, Tel Aviv 69978, Israel*

[*]To whom correspondence should be addressed. E-mail: bb8@columbia.edu (B.J.B), rabani@tau.ac.il (E.R), drr2103@columbia.edu (D.R.R)

**The very nature of glass is somewhat mysterious: while relaxation times in glasses are of sufficient magnitude that large-scale motion on the atomic level is essentially as slow as it is in the crystalline state, the structure of glass appears barely different than that of the liquid that produced it.[1,2,3,4] Quantum mechanical systems ranging from electron liquids to superfluid helium appear to form glasses, but as yet no unifying framework exists connecting classical and quantum regimes of vitrification. Here we develop new insights from theory and simulation into the quantum glass transition that surprisingly reveal distinct regions where quantum fluctuations can either promote or inhibit glass formation.**

While a wide variety of glassy systems ranging from metallic to colloidal can be accurately described using classical theory, quantum systems ranging from the electronic to magnetic appear to form glassy states.[5,6] Perhaps the most intriguing of these is that the coexistence of superfluidity and dynamical arrest, namely the "superglass" state. Recent numerical, theoretical and experimental work has given convincing evidence for the reality of this unusual arrested state of matter.[7,8,9] However, while such intriguing examples exist there

1

currently exists no unifying framework to treat the interplay between quantum and glassy fluctuations in the liquid state.

To attempt to formulate a theory for a quantum liquid to glass transition we may first appeal to the classical case for guidance. Here, a microscopic theory exists in the form of mode-coupling theory (MCT), which only requires simple static structural information as input and produces a full range of dynamical predictions for time correlation functions associated with single particle and collective fluctuations.[10] Although MCT has a propensity to overestimate a liquid's tendency to form a glass it has been shown to account for the emergence of the non-trivial growing dynamical length scales associated with vitrification.[11] Perhaps more importantly, MCT has made numerous non-trivial predictions ranging from logarithmic temporal decay of density fluctuations and reentrant dynamics in adhesive colloidal systems to various predictions concerning the effect of compositional mixing on glassy behaviour[12,13]. These have been confirmed by both simulation and experiment[14,15,16].

A fully microscopic quantum version of MCT (QMCT) that requires only the observable static structure factor as input may be developed along the same lines as the classical version. Indeed, a zero temperature version of such a theory has been developed and successfully describes the wave vector dependent dispersion in superfluid helium.[17] In the supplementary information, we outline the derivation of a full temperature dependent QMCT. In the limit of high temperatures, this theory reduces to the well-established classical MCT, while at zero temperature our theory reduces precisely to the aforementioned T=0 quantum theory. The structure of these two theories is dramatically different, suggesting the possibility of non-trivial emergent physics over the full range of parameters that tune between the classical and quantum limits.

The fully microscopic QMCT allows for a detailed description of the dynamical phase diagram that separates an ergodic fluid region from an arrested glassy one as a function of both thermodynamic control variables as well as the parameters (e.g. $\hbar$) that control the size of quantum fluctuations. To illustrate this, we perform detailed QMCT calculations on a hard-sphere system as a function of the system's volume fraction $\phi$ and the dimensionless parameter $\Lambda^*$ (the ratio of the de Broglie thermal wave length to the particle size) that controls the scale of quantum behaviour. Despite their simplicity, hard sphere systems are well characterized, experimentally realizable, and show all the features of glassy behaviour that are exhibited by more complex fluids. It is well-known from experiment and simulation that classical hard spheres enter a glassy regime for volume fractions in the range $\phi = 50 - 60\%$ independent of temperature.[18,19] Figure 1 shows the full structure of the dynamical phase diagram. The QMCT calculations are consistent with this in the classical limit ($\Lambda^* \to 0$), but upon departure from this show a rather remarkable reentrant behaviour. In particular, as the scale of quantum fluctuations are tuned from small to high values an initially flat regime is followed by the system becoming glassier and then finally favouring the fluid when quantum fluctuations are large. This behaviour is surprising given the fact that reentrance is not hinted at in the static structure factor. We also show the behaviour produced by a strictly classical MCT calculation performed with the quantum structure factor as input where only a featureless border separating liquid from glass is demonstrated. This fact clearly shows that the reentrant behaviour predicted by QMCT is a non-trivial product of the properties of the theory and not the static structure factor input.

To obtain a physical understanding of this surprising prediction we turn to the ring polymer molecular dynamics (RPMD) approach to quantum dynamics.[20] This method exploits the path integral formulation of quantum mechanics in which a quantum particle is mapped onto

3

a classical ring polymer consisting of a series of replicas linked by harmonic springs. Static properties can be calculated exactly using this mapping while RPMD utilizes the classical evolution of the polymers to provide an approximation to quantum dynamics. This approach has been previously shown to give accurate dynamical properties for systems ranging from nearly classical to those where tunnelling is dominant.[21,22]

We performed RPMD simulations for a binary Lennard-Jones system at a density and temperature that classically exhibits glassy behaviour (details provided in supplementary information).[23] Figure 2(a) shows the change in the diffusion coefficient of the particles as the quantum fluctuations of the system, controlled by varying $\Lambda^*$, are increased. This property shows the same reentrance seen in the QMCT results. The structure factor, shown in Figure 2(b), reveals only a monotonic broadening over the entire region under study.

Analysis of the RPMD trajectories allows us to deduce the origin of this effect. In Figure 2(c), we show the ratio of the average radius of gyration of the polymers representing each particle, a static property given exactly by the RPMD simulations, to its free particle value which is proportional to $\Lambda^*$. The spread of each polymer (or the width of the thermal wave packet) is directly related to the quantum mechanical uncertainty about its position. Hence, the uncertainty principle dictates that decreasing the width of the packet corresponds to an increase in kinetic energy.

The trend in the spread of the particles as shown in Figure 2(c) is in excellent agreement with that seen in the diffusion coefficient (see Figure 2(a)) and provides insight into the reentrant behaviour. As quantum fluctuations are introduced into the system the wave packet of each particle attempts to delocalize. Initially ($\Lambda^*$<0.1) thermally accessible space is available in the

system for the particle to expand into, allowing the radius of gyration to increase almost freely. The ratio of the spread of the particles to their free particle values is near unity and the diffusion is largely unchanged. However as $\Lambda^*$ is increased further, the width of the packet becomes comparable to size of the cage in which it resides. There is now little free space into which the packet may expand, leading to a dramatic decline in the ratio. Figure 3(a) shows a configuration typical of this regime in which the particle is localized in its cavity. This confinement in its position causes a large rise in the kinetic energy of localization exerting, as $\Lambda^*$ increases, a progressively larger quantum pressure on the cavity.

For diffusion to occur, the particles must rearrange in this highly crowded environment. This requires contraction of their wave packets as they pass through the narrow gaps, localizing them further and incurring an additional increase in their kinetic energy. This higher energy required to push through the gaps acts as a bottleneck to diffusion and leads to the slowing reflected in the intermediate $\Lambda^*$ region (see Figure 2(a)). This is also shown in the inset of Figure 3(a) which depicts the mean square displacement of the particle. A long intermediate beta relaxation regime is observed in which the particle is caged before being becoming mobile again at long times. When $\Lambda^*$ is raised further, the thermal wavelength becomes comparable to the particle size and the kinetic energy becomes sufficient to flood the barriers between cavities leading to a rise in the radius of gyration and the occurrence of tunnelling between the cavities, thereby facilitating diffusion. This can be seen in the representative snapshot shown in Figure 3(b) in which the particle is stretched across two cavities. Accordingly, the ratio of the radius of gyration to its free value recovers with a corresponding increase in diffusion and diminishing of the caging regime as shown in the inset of Figure 3(b).

The theory we have developed for the quantum glass transition predicts interesting generic dynamical anomalies such as a reentrant border between the disordered arrested and fluid regimes. Semi-classical quantum dynamics simulations display similar features and physically illuminate the origin of the predicted relaxation motifs. The physical interplay between crowding and quantum delocalization reported here, a generic feature of quantum glassy systems, might also be responsible for other physical phenomena. For example, it has been experimentally observed that lighter isotopes of hydrogen diffuse more slowly than heavier ones in water[24] and palladium[25] which has been recently elucidated by theory.[26] The regime where reentrance is observed is therefore realizable in chemical systems near ambient temperatures.

It is likely the reentrant transition observed here may also have implications beyond glassy systems. Intuition suggests that increasing quantum fluctuations monotonically enhances the exploration of the energy landscape. This forms the basis of the quantum annealing approach to optimization[27,28]. However, our work indicates that in certain regimes increasing quantum fluctuations can lead to dynamical arrest and hinder optimization. Indeed, reentrance has recently been observed in the dynamical phase behaviour of simple models of quantum optimization currently under investigation in the field of quantum information science.[29] Thus deep connections exist that unite these seemingly distinct physical systems and processes.

**Competing interests**

The authors declare that they have no competing financial interests.

**Figure Captions**

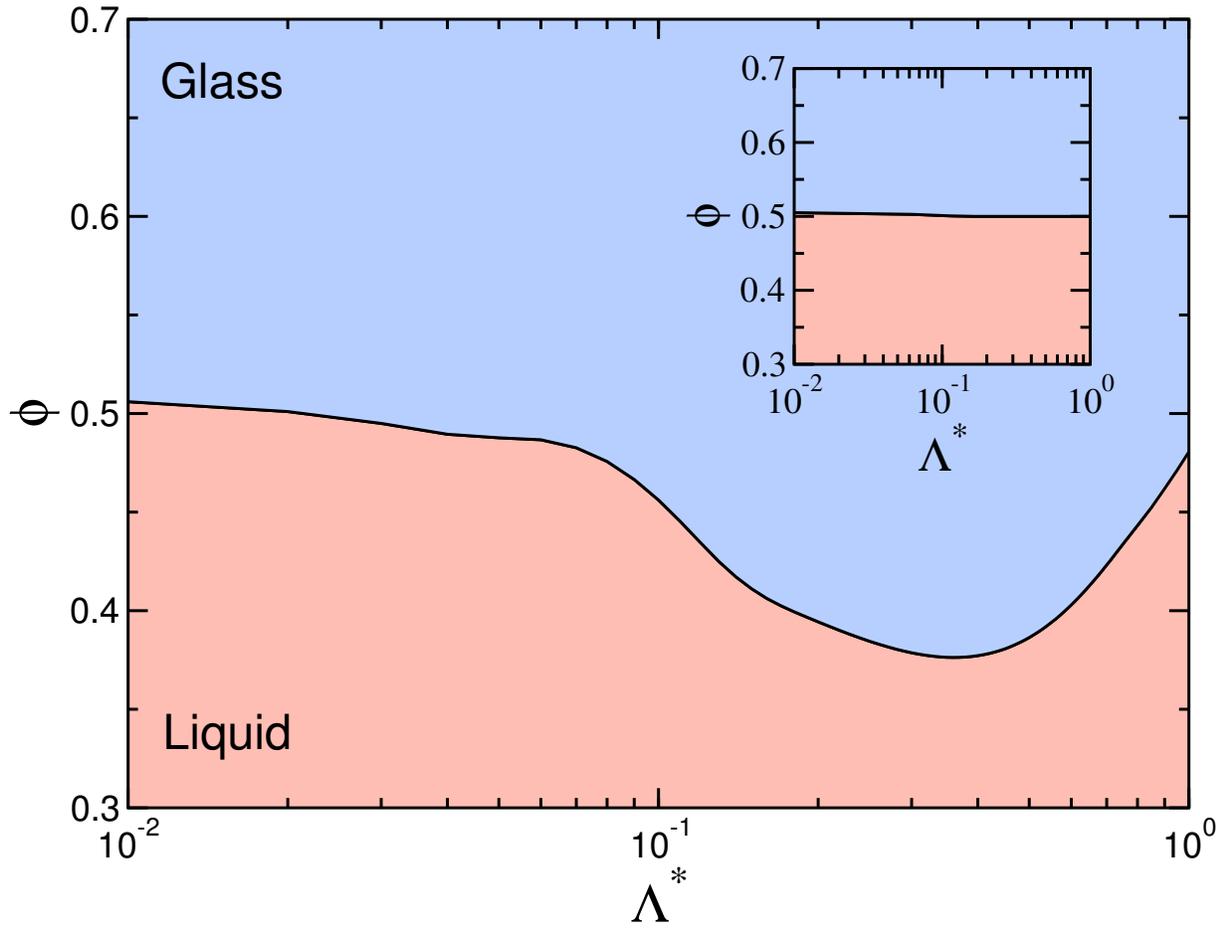

**Figure 1: Dynamic phase diagram calculated from the QMCT for a hard-sphere fluid.** $\phi$ is the volume fraction, $\Lambda^* = \sqrt{\beta\hbar^2/m\sigma^2}$ is the thermal wavelength in units of inter-particle separation $\sigma$, and $\beta = 1/k_B T$ is the inverse temperature. Inset: Dynamic phase diagram using the quantum mechanical input with a classical MCT.

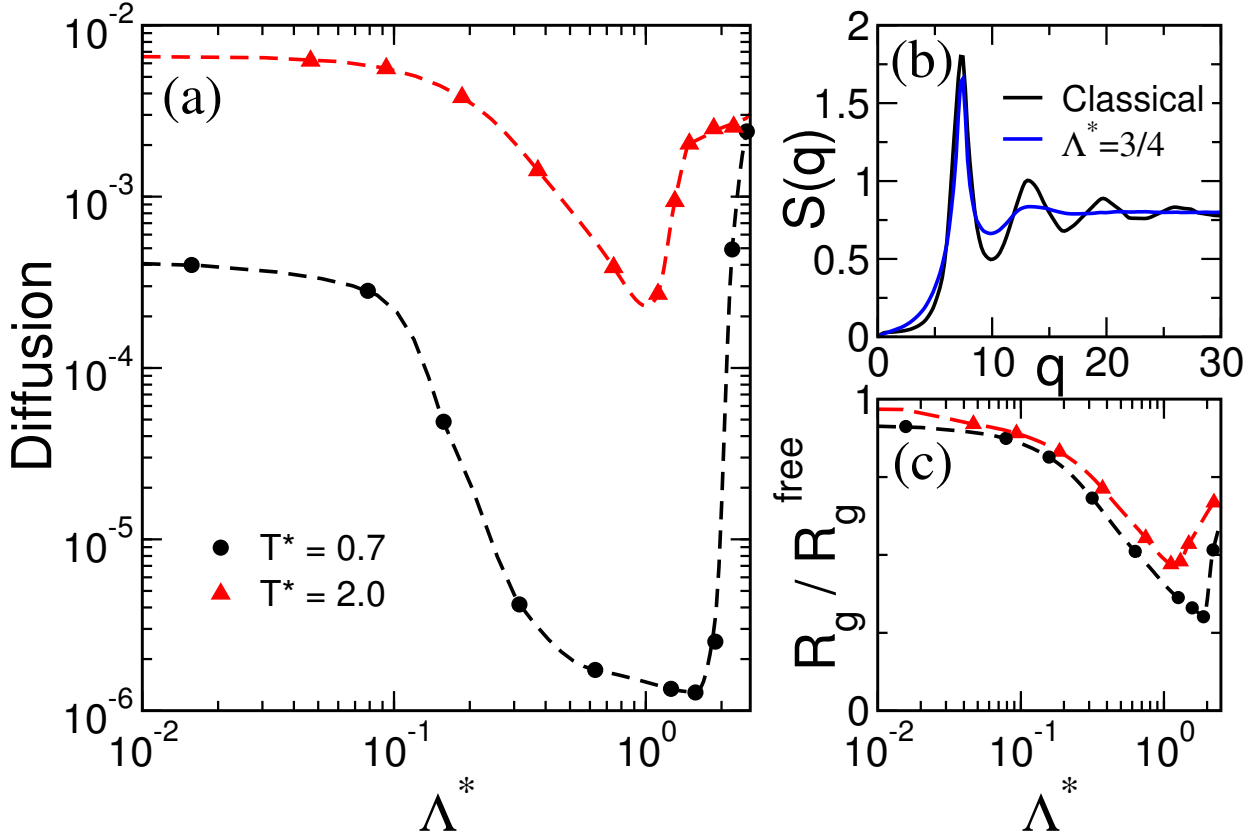

**Figure 2: Diffusion as function of quantumness from RPMD simulations.** (a) The diffusion constant as a function of the quantumness, $\Lambda^*$ obtained from the RPMD simulations for a quantum Kob-Anderson Lennard Jones binary mixture for $T^* = 2.0$ (red curve) and $T^*=0.7$ (black curve) . (b) Classical and quantum static structure factor of the "A" type particles. (c) Root-mean-square of the radius of gyration as a function of $\Lambda^*=0$ for the two systems shown in panel (a). The radius of gyration is defined as the average distance of the replicas from the polymer centre.

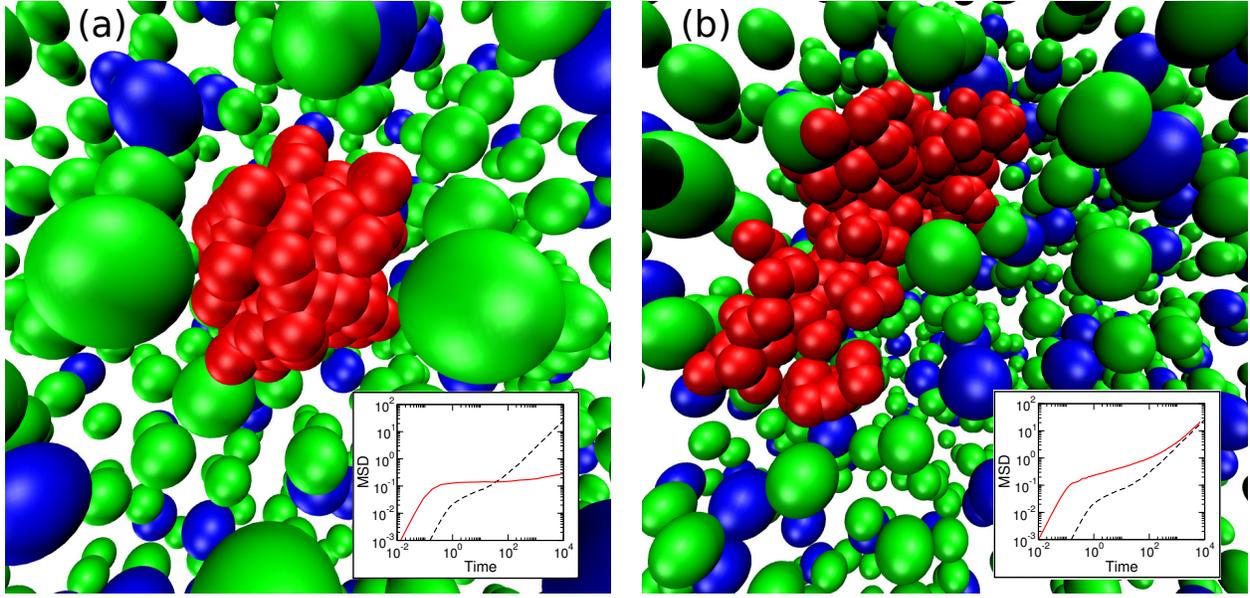

**Figure 3: Snapshots from the RPMD simulations taken from the caged ($\Lambda^*=0.7$, panel (a)) and tunnelling ($\Lambda^*=1.3$, panel (b)) regimes at $T^* = 0.7$.** For clarity all but one ring polymer in each snapshot is represented by its centre of mass. The red spheres represent the replicas of the polymer. Insets depict the mean square displacement, $\left\langle \left| \vec{R}(t) - \vec{R}(0) \right|^2 \right\rangle$, calculated from RPMD (solid curves) and classical MD (dashed curves) in Lennard Jones reduced units. The inset of panel (a) shows a long beta relaxation regime compared to the classical simulation and the tunnelling case shown in (b).



\newpage

\maketitle

\begin{affiliations}
 \item Department of Chemistry, Columbia University, 3000 Broadway, New York, New York, 10027, United States
 \item Institute of Physics, University of Tsukuba, Tsukuba, Japan
 \item School of Chemistry, The Sackler Faculty of Exact Sciences, Tel Aviv University, Tel Aviv 69978, Israel
\end{affiliations}

\section*{The quantum mode coupling theory}
We first outline the derivation and explicit expressions of the
quantum mode coupling theory (QMCT).  As discussed in the caption of
Fig.1, the magnitude of quantum fluctuations may be measured by a
dimensionless parameter that sets the ratio of the thermal wave
length, $\lambda$ to the particle size, $\sigma$ namely:
\begin{equation}
  \Lambda^{*}=\sqrt{\frac{\hbar^{2}}{k_{\rm B}T m \sigma^2}} = \frac{\lambda}{\sigma}
  \label{eq:lamstar}
\end{equation}
where $\hbar$ is Planck's constant divided by $2\pi$, $k_{\rm B}$ is
Boltzmann's constant, $T$ is the temperature, $m$ is the particle
mass, and $\sigma$ is the particle diameter.

Note that even for hard-spheres a temperature
appears that defines the scale of kinetic fluctuations. This may be
thought of as an intrinsic noise temperature following the quantum
fluctuation-dissipation theorem (QFDT).  Defining a projection
operator based on the Kubo transformation,
\begin{equation}
  (A|B) = \frac{1}{\hbar \beta}\int_{0}^{\hbar \beta}d\lambda\langle A(-i\lambda) B(0) \rangle = \frac{1}{\hbar \beta}\int_{0}^{\hbar \beta}d\lambda\langle A(0) B(i\lambda)\rangle,
  \label{eq:kubo_corr}
\end{equation}
where time evolution is defined via the standard Heisenberg picture:
\begin{equation}
A(t) = e^{i H t / \hbar} A e^{-i H t / \hbar},
\end{equation}
an exact quantum mechanical equation of motion is found for the Kubo-transform
of the density-density correlation function,
\begin{equation}
\ddot{\phi}_{q}(t) + \Omega_{q}^{2}\phi_{q}(t) +
\int_{0}^{t}dt'M_{q}(t')\dot{\phi}_{q}(t-t') = 0,
\label{eq:qgle}
\end{equation}
where $\Omega_{q}^{2}=\frac{k_{\rm B}T}{m\phi_{q}(0)}q^{2}$, $\phi_{q}(0)$
is the Kubo-transformed static structure factor, and $M_{q}(t)$ is the memory function.

At high temperature the frequency coefficient in the second term reduces to
\begin{equation}
\lim_{T\rightarrow \infty}
\Omega_{q}=\sqrt{\frac{k_{\rm B}Tq^{2}}{m S_{q}}},
\end{equation}
and at low temperature becomes,
\begin{equation}
\lim_{T\rightarrow 0}
\Omega_{q}=\frac{\hbar q^{2}}{2 m S_{q}} \equiv \omega_q
\end{equation}
which may be recognized as the well-known Bijl-Feynman dispersion at
zero-temperature.\cite{Bijl40,Feynman54}

Following the mode-coupling approach generalized to the quantum
mechanical context, the following expression for the memory function
of Eq.~(\ref{eq:qgle}) may be derived (in what follows we use the
notation $\tilde{C}(\omega)=\int_{\infty}^{\infty} dt \mbox{e}^{-i
  \omega t} C(t)$ for quantities in frequency space):
\begin{eqnarray}
\tilde{M}_{q}(\omega) &\approx& \frac{\hbar m \beta^{2}}{4\pi\omega
  q^{2} N} \sum_{{\bf k}} v_{q}^{2}(k,q-k)
\int_{-\infty}^{\infty}d\omega' \omega' \\ \nonumber & & \times
(\omega-\omega') T(\omega',\omega-\omega') \tilde{\phi}_{q-k}(\omega')
\tilde{\phi}_{k} (\omega-\omega'),
\label{eq:Mqmct}
\end{eqnarray}
where
\begin{eqnarray}
T(\omega_{1},\omega_{2})
= n(-\omega_{1}) n(-\omega_{2}) - n(\omega_{1}) n(\omega_{2}),
\label{eq:T}
\end{eqnarray}
and the vertex is given by
\begin{eqnarray}
v_{q}(k,q-k) &=& \frac{\Delta n(\Omega_{q-k})\Delta n(\Omega_{k})
  C_{q,k,q-k}}{S_{q-k}S_{k}K(\Omega_{q-k},\Omega_{k})}
\left[\frac{(\Omega_{k} +\Omega_{q-k})^{2}-\Omega_{q}^{2}}
  {(\Omega_{k}+\Omega_{q-k})}\right]
\label{eq:vertexGL}
\end{eqnarray}
with
\begin{eqnarray}
C_{q,k,q-k}&=&\frac{\Omega_{q}S_{q}S_{k}S_{q-k}-\frac{\Delta
n(\Omega_{q})}{2m}\left[q\cdot kS_{q-k}+q\cdot(q-k)S_{k}\right]}
{\Omega_{q}\Delta n(\Omega_{k}+\Omega_{q-k})
-(\Omega_{k}+\Omega_{q-k})\Delta n(\Omega_{q})}.
\label{eq:C}
\end{eqnarray}
Here $K(\Omega_{q-k},\Omega_{k}) = \frac{T(\Omega_{q-k},\Omega_{k})}
{\Omega_{q-k}+\Omega_{k}} + \frac{T(-\Omega_{q-k},\Omega_{k})}
{\Omega_{q-k}-\Omega_{k}}$, $\Delta n(\omega) = n(\omega)-n(-\omega)$
and $n(\omega)=\frac{1}{e^{\beta\hbar\omega}-1}$ is the Bose
distribution function at temperature $T$.

The above expressions close the equation of motion~\ref{eq:qgle} , and
require only the static structure factor to produce a full
approximation to the time dependence of the quantum density-density
time autocorrelation function.  To derive the full expressions quoted
above we have resorted to a finite temperature generalization of the
``resonance approximation'' of G\"{o}tze and
L\"{u}cke.\cite{Gotze76a,Gotze76b} However, the results presented in
Fig.~1 (and thus the predicted reentrance effect) are robust and do
not depend on this approximation. This can be demonstrated by
substituting a variety of different approximations to remove the
dependence of various static terms in the verticies on the
integrations over imaginary time induced by Kubo transformation.

Using the input of accurate quantum structure factors (as described in
the next subsection) one can make predictions as to the role of
quantum fluctuations on the glass transition. A version of QMCT has previously been developed to treat the quantum liquid regime.\cite{Rabanireview05} This theory is not capable of treating the regime where dynamics become glassy. A future article will detail the relationship between the theory used here and the previous version of QMCT.\cite{}

It may be shown analytically that the above equations reduce to the venerable
classical mode-coupling equations in the high temperature limit and to
the G{\"{o}}tze L\"{u}cke theory at $T=0$.  The latter theory produces
a representation of the dispersion of superfluid helium that is at
least as accurate as the Feynman-Cohen (FC) theory~\cite{Feynman56} at
low values of $q$ and exhibits Pitaevskii-bending of the spectrum at
high $q$, unlike the FC theory.  In particular at high $T$,
\begin{equation}
\lim_{\beta\rightarrow0}M_{q}(t)=\frac{k_{\rm B}T n}{16\pi^{3}m q^2}\int
d^{3}k\left({q}\cdot k c_{k}+ {q} \cdot(q-k)
c_{q-k}\right)^{2} \phi_{q-k}(t)\phi_{k}(t),
\label{eq:Mqtclassical}
\end{equation}
where $n$ is the number density and $c_{q} = \frac{1}{n}
\left(1-\frac{1}{S_{q}} \right)$ is the direct correlation function.
In addition, $\phi_{q}(t)$ reduces to the classical intermediate
scattering function, $F(q,t)$ as $\beta \to 0$. This is recognized as
the classical MCT memory function.\cite{Gotze2009}

At $T=0$ the equation for the memory function reduces to:
\begin{equation}
M_{q}(\omega)=\frac{m\beta^{2}}{2 n \omega q^{2}} \int
\frac{dk}{(2\pi)^{3}} v_{q}^{2}(k,q-k) \int_{0}^{\omega}
\frac{d\omega'}{\pi}\omega'(\omega-\omega')
\tilde{\phi}_{q-k}(\omega')\tilde{\phi}_{k}(\omega-\omega'),
\label{eq:MqtT=0}
\end{equation}
with
\begin{equation}
  v_{q}(k,q-k)=\frac{n}{2m}(\omega_{k}+\omega_{q-k}+\omega_{q})\left(q\cdot
  k c_{k}+ q\cdot(q-k)c_{q-k}\right)
\label{eq:vertexT=0}
\end{equation}
which are the $T=0$ equations for quantum density fluctuations in
superfluid helium first derived by G\"{o}tze and
L\"{u}cke.\cite{Gotze76a,Gotze76b} Note that in the $T=0$ case, the
entire structure of the memory function differs greatly from that of
its high temperature counterpart and the convolution structure is
lost.  Eqs.(\ref{eq:MqtT=0}) and (\ref{eq:vertexT=0}) do not imply a
memory function that is a product of correlators at identical times.
This is a consequence of the QFDT that must be satisfied. At $T=0$ the
function $T(\omega_q,\omega_k)$ becomes proportional to the difference
of a product of step-functions in frequency, dramatically altering
structure of the theory.  This distinction between the low and high
temperature limits has important consequences.  In addition to the
robust prediction of reentrance, we also find that glassy behavior
cannot be supported in the strict $T=0$ case.  Some or all of these
features seem to emerge both in certain quantum spin glasses and in
recent work on quantum versions of lattice models of glassy
liquids\cite{Zamponi}.  A future paper will be devoted to both a more
explicit derivation of the theory outlined here as well as the
physical implications of our work and the connection to other models
of quantum glass behavior.

\section*{Quantum integral equations for static structure}
The quantum integral equation approach used in this work to generate the input
required by the QMCT is based on the early work Chandler and
Richardson.\cite{Chandler84a,Chandler84b} For completeness, we provide
an outline of the approach. We begin with the Ornstein-Zernike
relation applicable to the quantum liquid. The quantum system composed
of $N$ particles can be mapped on a classical system consisting of $N$
ring polymers, each polymer being composed of $P$ beads. Then, we can
write the matrix RISM (reference interaction site
model~\cite{Chandler84a,Chandler84b}) equation for the classical
isomorphic system by:
\begin{equation}
h(|{\bf r}-{\bf r'}|) = \omega * c * \omega(|{\bf r}-{\bf r'}|) + n
\omega * c * h(|{\bf r}-{\bf r'}|),
\label{eq:QOZ}
\end{equation}
where $*$ denotes a convolution integral and $n$ is the number
density.  In the above equation, $h(r)$, $\omega(r)$, and $c(r)$ are
the total correlation function, the self correlation function, and
direct correlation function, respectively, defined by:
\begin{equation}
\begin{split} 
h(r) = \frac{1}{\hbar\beta}\int_0^{\hbar\beta} d\lambda h(r,\lambda) \\
\omega(r) = \frac{1}{\hbar\beta}\int_0^{\hbar\beta} d\lambda \omega(r,\lambda) \\
c(r) = \frac{1}{\hbar\beta}\int_0^{\hbar\beta} d\lambda c(r,\lambda) \\
\end{split} 
\label{eq:hwc}
\end{equation}
and $h(r,\lambda)$, $\omega(r,\lambda)$, and $c(r,\lambda)$ are the
imaginary time total, self, and direct correlation functions, respectively. In the
classical limit Eq.~(\ref{eq:QOZ}) reduces to the classical
Ornstein-Zernike equation with $\omega(r)=1$. In what follows, we will
use the notation $\tilde{\omega}_q(\lambda)$ for the Fourier transform
of $\omega(r,\lambda)$, and similarly for $\tilde{c}_q(\lambda)$ and
$\tilde{h}_q(\lambda)$:
\begin{equation}
\begin{split} 
\tilde{h_q} = \frac{1}{\hbar\beta}\int_0^{\hbar\beta} d\lambda
\tilde{h}_q(\lambda) \\ \tilde{\omega}_q =
\frac{1}{\hbar\beta}\int_0^{\hbar\beta} d\lambda
\tilde{\omega}_q(\lambda) \\ \tilde{c}_q =
\frac{1}{\hbar\beta}\int_0^{\hbar\beta} d\lambda \tilde{c}_q(\lambda)
\\
\end{split} 
\label{eq:hwcq}
\end{equation}

We now use the mean-pair interaction approximations along with the
quadratic reference action~\cite{Chandler84a} and rewrite:
\begin{equation}
  \tilde{\omega}_q(\lambda) = \exp\{ -q^2 R^2(\lambda)\},
\label{eq:wqlambda}
\end{equation}
where
\begin{equation}
  R^2(\lambda) = \sum_j \frac{1-\cos(\Omega_j \lambda)}{\beta m
    \Omega_j^2 + \alpha_j},
\label{eq:R}
\end{equation}
$m$ is the particle mass, $\Omega_j=2\pi j/\hbar \beta$ is the
Matsubara frequency and $\alpha_j$ is given by:
\begin{equation}
\alpha_j = \frac{1}{6 \pi^2 \hbar \beta} \int_0^{\infty} dq
\int_0^{\hbar \beta} d\lambda q^4 \tilde{v_q} (1-\cos(\Omega_j
\lambda) \tilde{\omega}(q,\lambda).
\label{eq:alpha}
\end{equation}
In the above the solvent induced self-interaction is given by:
\begin{equation}
\tilde{v_q} = -\tilde{c}_q^2(n \tilde{\omega}_q + n^2 \tilde{h}_q).
\label{eq:v}
\end{equation}

We now need to close the quantum Ornstein-Zernike equations, which in
$q$-space can be written as:
\begin{equation}
\tilde{h}_q = \tilde{\omega}_q \tilde{c}_q \tilde{\omega}_q + n
\tilde{\omega}_q \tilde{c}_q \tilde{h}_q.
\label{eq:QOZq}
\end{equation}
We use the Percus-Yevick closure of the form (in $r$-space):
\begin{equation}
c(r)=(h(r)+c(r)+1) (\exp(-\beta v(r)) - 1),
\label{eq:PY}
\end{equation}
where $v(r)$ is the pair interaction between two particles.
 
\section*{RPMD Simulations}
We performed RPMD simulations of the Kob-Andersen\cite{Kob95a}
binary Lennard-Jones (LJ) glass forming system. The Lennard Jones potential between particles $i$ and $j$ is given by,
\begin{equation}
V_{ij}(r_{ij}) = 4 \epsilon_{ij} \left[ \left( \frac{\sigma_{ij}}{r}\right)^{12}- \left(\frac{\sigma_{ij}}{r}\right)^{6} \right].
\end{equation}
The parameters and their conversion to atomic units as used in this work is given in Table \ref{tab:param}. The systems consisted of 1000 particles, 800 of type A and 200 of type B in a cubic box of length 9.4 $\sigma_{AA}$. The equations of motion were integrated using a timestep of 0.005 LJ units (0.35 fs) using the scheme of reference \cite{Ceriotti2010}. The simulations were carried out at constant volume for consistency with the QMCT results. The RPMD simulations were performed for distinguishable particles which is a valid approximation in the regime where the reentrance is observed.

The number of beads, $P$, used was given by the formula,
\begin{equation}
P = \frac{11.2 \hbar}{T^*}
\end{equation}
which was found to give good convergence for all the regimes studied.

Initial configurations were generated by annealing from a temperature
$T^{*}$=5.0 to the target temperature over a period of 1,000,000 timesteps. From these initial configurations we ran a further 200,000 steps of equilibration using the a targeted Langevin equation normal mode thermostatting scheme\cite{Ceriotti2010}. This
was followed by microcanonical dynamics for 2,000,000 steps during which the results
were collected. The quantum effect, $\Lambda^{*}$, was varied by changing the parameter $\hbar$. 
Five simulations were run for each temperature and value of $\hbar$ and the results averaged.

The root mean square radius of gyration is defined as,
\begin{equation}
r_i^{G} = \left<\frac{1}{P}\sum_{k=1}^P |{\bf r}_i^{(k)}-{\bf
  r}_i^{(c)}|^2\right>^{1/2},
\end{equation}
where
\begin{equation}
{\bf r}_{i}^{(c)}=\frac{1}{P} \sum_{k=1}^{P} {\bf r}_{i}^{(k)}
\end{equation}
is the center of mass of the ring polymer representing particle $i$. In the free limit the radius of gyration is \cite{Miller2005},
\begin{equation}
r^{G}_{i} = \frac{1}{2} \sqrt{\frac{\hbar^{2}}{k_{\rm B} T m_{i}}} = \lambda/2
\end{equation}
which is related to the De Broglie thermal wavelength as defined in Eq. \ref{eq:lamstar} via multiplication by two.

\begin{table}
\begin{tabular*}{0.5\textwidth}{@{\extracolsep{\fill}}ccc} 
\hline\hline 
\multicolumn{1}{c}{Parameter} & \multicolumn{1}{c}{LJ units} & \multicolumn{1}{c}{Atomic Units} \\ 
\hline
$\epsilon_{AA}$ & 1  &  3.8x10$^{-4}$\\
$\epsilon_{BB}$ & 0.5  &  1.9x10$^{-4}$\\
$\epsilon_{AB}$ & 1.5  &  5.7x10$^{-4}$\\
$\sigma_{AA}$   & 1  & 6.43 \\
$\sigma_{BB}$   & 0.88  & 5.65 \\
$\sigma_{AB}$   & 0.8  & 5.14 \\
Mass$_{A}$      & 1  & 3646 \\ 
Mass$_{B}$      & 1  & 3646 \\\hline \hline
\end{tabular*}
\caption[]{Parameters used in our RPMD simulations on the Andersen-Kob Lennard-Jones glass forming system.}  
\label{tab:param}
\end{table}

\pagebreak

\end{document}